\documentclass[prb,twocolumn,superscriptaddress,showpacs,amsmath,amssymb]{revtex4}
\usepackage{float}
\usepackage{amsmath}
\usepackage{graphicx}
\usepackage{hyperref}
\usepackage{bm}
\usepackage[usenames]{color}
\usepackage{verbatim}
\newcommand{\be}{\begin{equation}}
\newcommand{\ee}{\end{equation}}   
\newcommand{\bea}{\begin{eqnarray}}
\newcommand{\eea}{\end{eqnarray}}
\newcommand{\phrl}[1]{Phys.~Rev.~Lett. {\bf #1}}
\newcommand{\phrb}[1]{Phys.~Rev.~B {\bf #1}}
\newcommand{\phrx}[1]{Phys.~Rev.~X {\bf #1}}
\newcommand{\cmat}[1]{arXiv:{\bf #1}}

\newcommand{\jpcm}[1]{J.~Phys.:Condens.~Matter.{\bf #1}}
\newcommand{\bib}{\bibitem}
\newcommand{\lb}{\left[}
\newcommand{\rb}{\right]}
\newcommand{\lp}{\left(}
\newcommand{\rp}{\right)}
\newcommand{\q}{\mathbf{q}}
\renewcommand{\k}{\mathbf{k}}
\newcommand{\tk}{\tilde{k}_{s'}}
\newcommand{\tkz}{\tilde{k}_{z,s'}}

\begin{document}

\title{Absorption of circular polarized light in tilted Type-I and II Weyl semimetals}

\author{S. P. Mukherjee}
\affiliation{Department of Physics and Astronomy, McMaster University, Hamiltion, Ontario, Canada L8S 4M1}

\author{J. P. Carbotte}
\affiliation{Department of Physics and Astronomy, McMaster University, Hamiltion, Ontario, Canada L8S 4M1}
\affiliation{Canadian Institute for Advanced Research, Toronto, Ontario, Canada M5G 1Z8}

\begin{abstract}
We calculate the A.C. optical response to circularly polarized light of a Weyl semimetal (WSM) with varying amounts of tilt of the Dirac cones. Both type-I and II 
(overtilted) WSM are considered in a continuum model with broken time reversal (TR) symmetry. The Weyl nodes appear in pairs of equal energies but of opposite momentum 
and chirality. For type-I the response of a particular node to right (RHP) and left (LHP) hand polarized light are distinct only in a limited range of photon energy 
$\Omega$, $\frac{2}{1+C_{2}/v}<\frac{\Omega}{\mu}<\frac{2}{1-C_{2}/v}$ with $\mu$ the chemical potential and $C_{2}$ the tilt associated with the positive chirality node 
assuming the two nodes are oppositely tilted. For the over tilted case (type-II) the same lower bound applies but there is no upper bound. If the tilt is reversed the 
RHP and LHP response are also reversed. We present corresponding results for the Hall angle.

\end{abstract}


\maketitle

\section{Introduction}
\label{sec:I}

A number of new materials have been found to be Weyl semimetals with pairs of Weyl nodes displaying opposite chirality. Among these are TaAs \cite{Lv,Xu,Ding,Shi,Yang}, 
NbAs \cite{Alidoust}, YbMnBi$_{2}$ \cite{Borisenko}, pyrochlore iridates \cite{Wan} and HgCr$_2$Se$_4$ \cite{Fang}. These materials exhibit exotic properties such as 
surface states with Fermi arcs \cite{Potter, Moll} and negative magnetoresistance \cite{Kim,Huang} associated with the chiral anomaly. They also exhibit an anomalous Hall 
effect \cite{Burkov, Zyuzin,Balents, Tiwari, Steiner}. The longitudinal dynamic optical conductivity which gives direct and valuable information on the dynamics of the 
charge carriers has been experimentally investigated in a number of Dirac and Weyl semimetals \cite{Chen, Sushkov, Xiao, Neubauer} for which a linear in photon energy
interband background is expected \cite{Ashby, Timusk}. This linear dependence reflects the 3-dimensional (3D) nature of the energy bands as well as the linearity of the 
dispersion curves. For graphene which is 2-dimensional the interband background is instead constant \cite{Jiang, Gusynin}. Deviations from these simple laws can arise 
for more complicated band structures \cite{Koshino, Nicol, Tabert, Li} and from correlation effects \cite{Sharapov, Stauber, Peres, Kuzmenko, LeBlanc, Carbotte} and these 
provide additional important information. In a recent optical study  \cite{Chinotti} in YbMnBi$_{2}$ two quasilinear energy regions are identified as expected in the
theoretical \cite{Tabert} model of the broken time reversal symmetry of references [\onlinecite{Koshino}] and  [\onlinecite{Nicol}]. In the Dirac semimetal 
Cd$_{3}$As$_{2}$ \cite{Neubauer} the interband background is observed to vary with photon energy $\Omega$ as $\Omega^{z'}$ where the exponent $z'=1.65$ which can be 
identified with a sublinear $\epsilon(\k)=|\k|^{z} (z=0.6)$ electron dispersion as shown by B$\acute{a}$csi and Virosztek [\onlinecite{Virosztek}] who derived the 
relationship $z'=\frac{D-2}{z}$ with $D$ the dimension, here equal to 3.

The Dirac cones in a WSM which define the charge carriers dispersion curves can be tilted away from the vertical axis. A WSM can be classified as type-I or type-II 
depending on the degree of tilt. For type-I the tilt is assumed to be smaller than the Fermi velocity $v$ and for the undopped case the Fermi surface is a single point 
consistent with the Weyl node. When the tilt (overtilted case) becomes larger than $v$ the Fermi surface is no longer just a point. There exists a hole and an 
electron pocket and the density of states at the Fermi surface is finite. This is referred to as type-II WSM. For a WSM with broken time reversal symmetry the Weyl nodes 
come in pairs of equal energy but displaced in momentum from each other and their chirality is opposite. If, in addition, inversion symmetry is broken the Weyl points 
are no longer  at the same energy. The effect of a tilt on the dynamical longitudinal optical conductivity was studied by Carbotte [\onlinecite{Carbotte1}] in the case 
of broken TR invariance. It was found that, for a given value of the chemical potential $\mu$, the expected linear law in photon energy $\Omega$ remained for 
$\frac{\Omega}{\mu}>\frac{2}{(1-C_{2}/v)}$ for type-I with $\frac{C_{2}}{v}<1$. In the range $\frac{2}{(1+C_{2}/v)}$ to $\frac{2}{(1-C_{2}/v)}$ there are characteristic 
modifications related to the amount of tilt involved. Below $\frac{\Omega}{\mu}=\frac{2}{(1+C_{2}/v)}$ the longitudinal optical response is zero. This is to 
be contrasted to the case when the tilt is zero for which we get zero up to $2\mu$ and an unmodified linear law above. For type-II with the tilt $\frac{C_{2}}{v}>1$ 
modifications to the linear law persist to high value of $\Omega$. These again start at $\frac{\Omega}{\mu}=\frac{2}{(1+C_{2}/v)}$ below which the 
conductivity is zero. In a very recent preprint Steiner et. al [\onlinecite{Steiner}] have given results for the A.C. Hall conductivity in the case of type-I WSM and 
find in our notation that it is non zero only in a confined photon energy range $\frac{2}{1+\frac{C_{2}}{v}}<\frac{\Omega}{\mu}<\frac{2}{1-\frac{C_{2}}{v}}$.

In this paper we consider the effect of a tilt on the absorption of circular polarized light. We consider both the case of type I and II. Right and left handed 
conductivity $\sigma_{+}(T=0,\Omega)$ and $\sigma_{-}(T=0,\Omega)$ are calculated as is the related Hall angle. In section-II we specify the basic continuum model Hamiltonian 
on which all our calculations are based. The Green's function underlying this model is specified and used in a Kubo formula at zero temperature $(T=0)$ to obtain the 
anomalous Hall conductivity $\sigma_{xy}(T=0,\Omega)$. For the real part of $\sigma_{xy}(T=0,\Omega)$ in the D.C. limit we recover the results of Ref.[\onlinecite{Tiwari}] 
and for the imaginary part at finite photon energy we recover the results of Ref.[\onlinecite{Steiner}] in the case when the tilt $\frac{C}{v}$ is less than one. New analytic 
results are established in the overtilted case and these are compared graphically with the $\frac{C}{v}<1$ case. In section-III we construct from the absorptive (imaginary) 
part of the Hall conductivity $\Im\sigma_{xy}(T=0,\Omega)$ and results for the real part of the longitudinal conductivity \cite{Carbotte1} (absorptive part) 
$\Re\sigma_{xx}(T=0,\Omega)$, the conductivities $\sigma_{+}(T=0,\Omega)$ and $\sigma_{-}(T=0,\Omega)$ which describe the absorption RHP and LHP light respectively. In 
section-IV we discuss the Hall angle associated with polarized light and in section-V we provide further discussion and state our conclusions.

\section{Formalism and Hall conductivity}
\label{sec:II}

Following the notation of Ref.[\onlinecite{Tiwari}] we start with the simplest continuum Hamiltonian for pair of Weyl nodes denoted by 1 and 2 of opposite chirality at 
$k_{z}\mp Q$ along the $z$-axis with tilt $C_{1},C_{2}$ and Fermi velocity $v$.
\bea
&& \hat{H}_{1,2}(\k)=C_{1,2}(k_{z}\mp Q)\pm v \bm{\sigma}.(\k\mp Q\bm{e}_z)\nonumber\\
&& =C_{1,2}(k_{z}-s' Q)+s' v \bm{\sigma}.(\k-s'Q\bm{e}_z)
\label{Hamiltonian}
\eea
where $s'=1$ for Weyl point indexed by 1 and  $s'=-1$ for Weyl point indexed by 2. $\bm{e}_{i}$ is the unit vector along the axis $x_{i}$ where $i=x, y, z$. The Pauli 
matrices are defined as usually by,
\be
\sigma_{x}=\lp\begin{array}{cc}0 & 1\\ 1 & 0 \end{array} \rp, \sigma_{y}=\lp\begin{array}{cc}0 & -\imath\\ \imath & 0 \end{array} \rp, \sigma_{z}=\lp\begin{array}{cc}1 & 0\\ 0 & -1 \end{array} \rp,
\ee
We define the new variable $\tkz=k_{z}-s'Q$. The Green's function corresponding to the above Hamiltonian is given by,
\be
G_{s'}(k,z)= \lb I_{2}z-\hat{H}_{s'}(\k)\rb^{-1},
\label{GF-definition}
\ee
where $I_{2}$ is a $2\times 2$ unit matrix. It is straight forward to show that one can write Eq.(\ref{GF-definition}) explicitly in matrix form as,
\bea
&& G_{s'}(k,z)= -\frac{1}{2v\tk}\sum_{s=\pm} \frac{s}{z-C_{s'}\tkz+sv\tk} \times \nonumber \\
&& \lp \begin{array}{cc}z-(C_{s'}-s'v)\tkz & s'v(k_{x} -\imath k_{y})\\ s'v(k_{x} +\imath k_{y}) & z-(C_{s'}+s'v)\tkz \end{array} \rp,
\label{GF-matrix-form}
\eea
where we have introduced the new symbol $\tk=\sqrt{k^2_{x}+k^2_{y}+\tkz^2}=|\k-s'Q\bm{e}_z|$. Following standard algebra we can write the full Green's function as,
\bea
&& G_{s'}(k,z)= \frac{1}{2}\sum_{s=\pm} \frac{1}{z-C_{s'}\tkz+sv\tk} \times \nonumber\\
&& \lp \begin{array}{cc}1-ss'\lp\tkz/\tk\rp & -ss'\{(k_{x} -\imath k_{y})/\tk\}\\ -ss'\{(k_{x} +\imath k_{y})/\tk\} &  1+ss'\lp\tkz/\tk\rp\end{array} \rp, \nonumber
\eea
which when written following the notation in Ref.[\onlinecite{Tiwari}] is,
\be
G_{1,2}(k,\imath \omega_{n})= \sum_{s=\pm} \frac{1-ss'\bm{\sigma}.\bm{N}_{k\mp Q\hat{e}_z}}{\imath \omega_{n}-C_{1,2}(k_{z}\mp Q)+sv|\k\mp Q\bm{e}_z|},
\label{GF-full}
\ee
where $\bm{N}_{k\mp Q\hat{e}_z}=\frac{k_{x}\bm{e}_{x}+k_{y}\bm{e}_{y}+(k_{z}\mp Q)\bm{e}_{z}}{\sqrt{k^2_{x}+k^2_{y}+\tkz^2}}$.

The current-current correlation function associated with the $xy$ component of the Hall conductivity is defined as,
\bea
&& \Pi_{xy}(\Omega,\q)=T\sum_{\omega_{n}}\sum_{s'=\pm} \int \frac{d^3k}{(2\pi)^3} \times \nonumber \\
&& J_{x,s'}G_{s'}(\k+\q,\omega_{n}+\Omega_{m})\times J_{y,s'}G_{s'}(\k,\omega_{n})\nonumber\\
&& =Te^2v^2\sum_{\omega_{n}}\sum_{s'=\pm} \int \frac{d^3k}{(2\pi)^3} \times \nonumber\\ 
&& \sigma_{x}G_{s'}(\k+\q,\omega_{n}+\Omega_{m})\times \sigma_{y}G_{s'}(\k,\omega_{n}),
\label{Pi(xy)}
\eea
where the current operators are,
\be
J_{\{x,y\},s'}=s'ev\sigma_{\{x,y\}}.
\label{Current-operator}
\ee

The dynamic Hall conductivity $\sigma_{xy}(T,\Omega)$ is given in terms of the off-diagonal current-current correlation function $\Pi_{xy},$

\bea
&& \sigma_{xy}(T,\Omega)=- \frac{\Pi_{xy}(\Omega,0)}{\imath\Omega} \nonumber\\
&& =-\frac{e^2}{\imath \Omega} \sum_{s'=\pm}s'\int^{\Lambda-s'Q}_{-\Lambda-s'Q} \frac{dk_{z}}{2\pi}\int^{\infty}_{0} \frac{k_{\perp}dk_{\perp}}{2\pi} \times \nonumber\\
&& \{f(C_{s'}k_{z}+vk)-f(C_{s'}k_{z}-vk)\} 2v^2\Omega \frac{k_{z}}{k}\times \nonumber\\
&& \lb \pi \delta(4v^2k^2-\Omega^2)-\frac{\imath}{4v^2k^2-\Omega^2}\rb \nonumber\\
&& =\frac{e^2 v^2}{2\pi^2} \sum_{s'=\pm}s'\int^{\Lambda-s'Q}_{-\Lambda-s'Q} k_{z} dk_{z} \int^{\infty}_{0} \frac{k_{\perp}dk_{\perp}}{k} \times \nonumber \\
&& \{f(C_{s'}k_{z}+vk)-f(C_{s'}k_{z}-vk)\} \times \nonumber \\
&& \lb\frac{1}{4v^2k^2-\Omega^2}+\imath\pi \delta(4v^2k^2-\Omega^2)\rb.
\label{sigmaXY}
\eea
with $f$ the Fermi-Dirac distribution at temperature $T$.
Here we have introduced a large cut off $\Lambda$ on the $k_{z}$-axis. Also since $k=\sqrt{k^2_{\perp}+k^{2}_{z}}$ we can replace the integration variable $k_{\perp}$ 
by $k$ (treating $k_{z}$ as constant). The real part of the D.C. transverse conductivity $\Re {\sigma_{xy}}$ is,
\bea
&& \Re {\sigma_{xy}}(T,\Omega=0)=\frac{e^2}{8\pi^2} \sum_{s'=\pm}s'\int^{\Lambda-s'Q}_{-\Lambda-s'Q} k_{z} dk_{z} \int^{\infty}_{0} dk \nonumber\\
&& \{f(C_{s'}k_{z}+vk)-f(C_{s'}k_{z}-vk)\} \frac{1}{k^2},
\eea
which can be reduced to the known result (Eq.(8) of Ref.[\onlinecite{Tiwari}]) namely,
\be
\Re\sigma_{xy}(T=0,\Omega=0)=\frac{e^2Q}{4\pi^2} \sum_{s'=\pm} \text{min}\lb 1,\frac{v}{|c_{s'}|}\rb.
\label{DCsigmaXY}
\ee

\begin{figure}[H]
\vspace{-2.8cm}
\hspace{-3.5cm}
\includegraphics[width=4.0in,height=6.2in, angle=270]{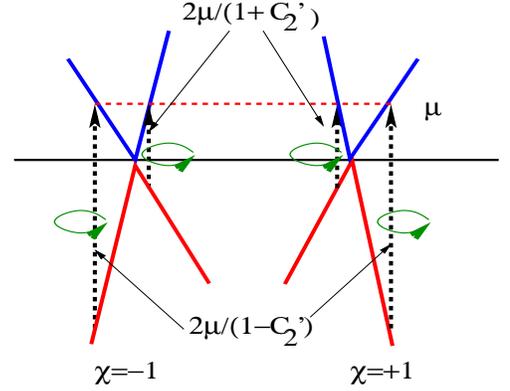} 
\vspace{-2.5cm}
\caption{(Color online) Here we scematically show two oppositely tilted Weyl cones which corresponds to the case $C_{1}=-C_{2}$ and for $0<C'_{2}<1$. We also show with 
black vertical arrows the limiting transitions possible for the tilted case with a specific chemical potential $\mu$ for circularly polarized light of photon energy 
$\Omega$.} 
\label{Cartoon}
\end{figure}

Returning to Eq.(\ref{sigmaXY}) for the dynamic Hall conductivity at finite $\Omega$ and taking its imaginary part we get,
\bea
&& \Im\sigma_{xy}(T,\Omega)=\frac{e^2v^2}{2\pi} \sum_{s'=\pm}s'\int^{\Lambda-s'Q}_{-\Lambda-s'Q} k_z dk_{z}\int^{\infty}_{0} \frac{k_{\perp}dk_{\perp}}{k} \times \nonumber\\
&& \{f(C_{s'}k_{z}+vk)-f(C_{s'}k_{z}-vk)\} \delta(4v^2k^2-\Omega^2)
\label{Im-ACsigmaXY}
\eea
We use the following property of Dirac delta function to write the imaginary part in simpler form,
\be
\delta(f(x))=\sum_{x_{i}}\frac{\delta(x-x_{i})}{|f'(x_{i})|} 
\ee
where $x_{i}$'s are the zeros of the function $f(x)$. We substitute this in the expression for $\Im\sigma_{xy}(\Omega)$ in Eq.(\ref{Im-ACsigmaXY}) to get,
\bea
&& \Im\sigma_{xy}(T,\Omega)=\frac{e^2v}{8\pi\Omega} \sum_{s'=\pm}s'\int^{\Lambda-s'Q}_{-\Lambda-s'Q} k_z dk_{z}\times \nonumber\\
&& \{f(C_{s'}k_{z}+\frac{\Omega}{2})-f(C_{s'}k_{z}-\frac{\Omega}{2})\} \lb 1-\Theta(|k_{z}|-\frac{\Omega}{2v})\rb\nonumber\\
&&\hspace{-0.5cm} =\frac{e^2v}{8\pi\Omega}\hspace{-0.2cm} \sum_{s'=\pm}\hspace{-0.2cm} s'\hspace{-0.2cm} \int^{\frac{\Omega}{2v}}_{-\frac{\Omega}{2v}} \hspace{-0.2cm} k_z \{f(C_{s'}k_{z}+\frac{\Omega}{2})-f(C_{s'}k_{z}-\frac{\Omega}{2})\} dk_{z}.\nonumber\\
\eea
Here we have also changed the variable $k_{\perp}$ to $k$ as was described for the D.C. case. At this point we see that when $C_{1}=C_{2}$ i.e. when both the cones are 
tilted in the same direction then $\Im(\sigma_{xy}(\Omega))$ is identically zero. On the other hand, when $C_{1}=-C_{2}$ (oppositely tilted case), (see Fig.(\ref{Cartoon}))
which means making the replacement $C_{s'}=-s'C_{2}$ in the above equation we get instead,
\bea
&& \Im\sigma_{xy}(T,\Omega) =\frac{e^2v}{8\pi\Omega} \sum_{s'=\pm}s'\int^{\frac{\Omega}{2v}}_{-\frac{\Omega}{2v}} dk_z k_{z} \times \nonumber\\
&& \{f(-s'C_{2}k_{z}+\frac{\Omega}{2})-f(-s'C_{2}k_{z}-\frac{\Omega}{2})\} \nonumber\\
&& \hspace{-0.8cm}=\frac{e^2v}{8\pi\Omega} \hspace{-0.2cm}\sum_{s'=\pm}\hspace{-0.3cm} s'\hspace{-0.2cm} \int^{\hspace{-0.1cm}-s'\frac{\Omega}{2v}}_{\hspace{-0.05cm}s'\frac{\Omega}{2v}}\hspace{-0.4cm}  k_z \{f(C_{2}k_{z}+\frac{\Omega}{2})-f(C_{2}k_{z}-\frac{\Omega}{2})\} dk_{z},\nonumber\\
\eea
here we have replaced $s'k_{z}$ by $k_{z}$.
\bea
&& \Im\sigma_{xy}(T,\Omega)=-\frac{e^2v}{8\pi\Omega} \sum_{s'=\pm}s'\int^{s'\frac{\Omega}{2v}}_{-s'\frac{\Omega}{2v}}dk_{z} k_z \times \nonumber \\
&& \{f(C_{2}k_{z}+\frac{\Omega}{2})-f(C_{2}k_{z}-\frac{\Omega}{2})\} dk_{z}\nonumber\\
&& \hspace{-0.8cm}= -\frac{e^2v}{4\pi\Omega} \int^{\frac{\Omega}{2v}}_{-\frac{\Omega}{2v}}k_z \{f(C_{2}k_{z}+\frac{\Omega}{2})-f(C_{2}k_{z}-\frac{\Omega}{2})\} dk_{z}.\nonumber\\
\label{Im-ACsigmaXY-opp-tilt}
\eea
Now we take the limit of temperature $T$ going to zero and replace the Fermi function by Heaviside step function $\Theta$ as shown below.
\bea 
&& \lim_{T \to 0}\{f(C_{2}k_{z}+\frac{\Omega}{2})-f(C_{2}k_{z}-\frac{\Omega}{2})\} \nonumber \\
&& =\Theta(-C_{2}k_{z}-\frac{\Omega}{2}+\mu)-\Theta(-C_{2}k_{z}+\frac{\Omega}{2}+\mu) \nonumber\\
&& =\Theta(C_{2}k_{z}-\frac{\Omega}{2}-\mu)-\Theta(C_{2}k_{z}+\frac{\Omega}{2}-\mu)
\eea
which gives,
\bea
&&\hspace{-0.4cm}\Im\sigma_{xy}(T=0,\Omega)= -\frac{e^2v}{4\pi\Omega} \int^{\frac{\Omega}{2v}}_{0} \hspace{-0.2cm} dk_{z} k_z \bigl[ \Theta(C_{2}k_{z}-\frac{\Omega}{2}-\mu)-\nonumber\\
&& \Theta(C_{2}k_{z}+\frac{\Omega}{2}-\mu) +\Theta(-C_{2}k_{z}+ \frac{\Omega}{2}-\mu)\bigr].
\label{Im-ACsigmaXY-Tzero}
\eea
We see that simplifications can be made to Eq.(\ref{Im-ACsigmaXY-Tzero}) depending on the relative magnitude of the chemical potential $\mu$ and the photon energy
$\Omega$. For $\mu>\frac{\Omega}{2}$ the third theta function drops out as its argument becomes negative under this condition. On the contrary when 
$\frac{\Omega}{2}>\mu$ the second theta function in the square bracket always produces one. Considering these together with the conditions $0<C'_{2}<1$ or $C'_{2}>1$ we 
arrive at the results which we summarize below. To state our results we have assumed that for any variable $a$, $a'=a/v$.

For $0<C'_{2}<1$ which corresponds to the WSM type-I we get only a finite region in $\Omega$ within which the imaginary part of the anomalous Hall conductivity 
$\Im\sigma_{xy}(T=0,\Omega)$ is non-zero. Namely 
\bea
&& \frac{\Im\sigma_{xy}(T=0,\Omega)}{\mu'e^2/8\pi}=0,~~\text{for}~~~~ \tilde{\Omega}<\frac{2}{1+C'_{2}}\nonumber\\
&& =\lb\frac{1}{4}(1-\frac{1}{C'^2_{2}})\tilde{\Omega}+\frac{1}{C'^2_{2}}-\frac{1}{C'^2_{2}}\frac{1}{\tilde{\Omega}}\rb,\nonumber \\
&& ~~~~~~~~~~~~~~~~~~~~~~~\text{for}~\frac{2}{1-C'_{2}}>\tilde{\Omega}>\frac{2}{1+C'_{2}}\nonumber\\
&& =0,\hspace{2.0cm}\text{for}~\tilde{\Omega}>\frac{2}{1-C'_{2}}
\label{Im-ACsigmaXY-0<c2<1}
\eea
Here $\tilde{\Omega}=\Omega'/\mu'=\Omega/\mu$. This agrees with Ref.[\onlinecite{Steiner}] when the change in notation is accounted for. The limits $\frac{2}{1+C'_{2}}$
and $\frac{2}{1+C'_{2}}$ are identified as the onsets of possible interband optical transitions in Fig.(\ref{Cartoon}) including a tilt $C'_{2}<1$.
\begin{figure}[H]
\begin{center}
\includegraphics[width=2.5in,height=3.2in, angle=270]{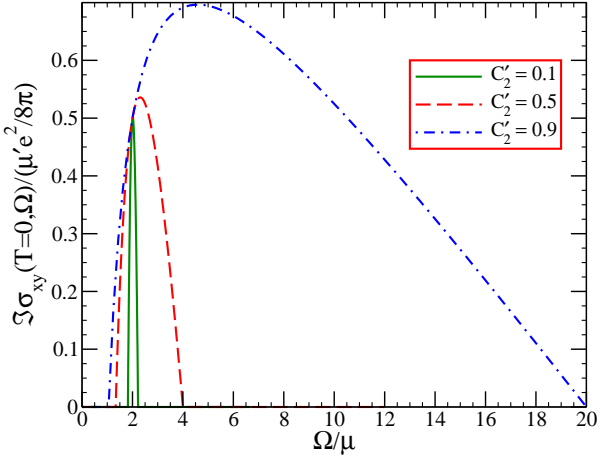} 
\vspace{0.5cm}
\caption{(Color online) Imaginary anomalous Hall conductivity $\Im\sigma_{xy}(T=0,\Omega)$ in units of $\frac{\mu' e^2}{8\pi}$ (where $\mu'=\mu/v$) is plotted against the 
photon energy 
$\Omega$ normalized by $\mu$ for three different values of the tilt parameter $C'_{2}$ which represent the type-I WSM. For all $C'_{2}$ we see the dome like structures 
as described in Eq.(\ref{Im-ACsigmaXY-0<c2<1}) in the range $\frac{2}{1+C'_{2}}<\frac{\Omega}{\mu}<\frac{2}{1-C'_{2}}$. For photon energies outside this range 
$\Im\sigma_{xy}(T=0,\Omega)$ becomes zero.}
\label{xy-0<c2<1}
\end{center}
\end{figure}

For the overtilted case satisfying the condition $C'_{2}>1$ which corresponds to WSM type-II, we get two distinct regions in $\Omega$ where the imaginary part of the 
anomalous Hall conductivity $\Im\sigma_{xy}(T=0,\Omega)$ is non-zero.
\bea
&& \frac{\Im\sigma_{xy}(T=0,\Omega)}{\mu'e^2/8\pi}=0,~~\text{for}~~~~ \tilde{\Omega}<\frac{2}{1+C'_{2}}\nonumber\\
&& =\lb\frac{1}{4}(1-\frac{1}{C'^2_{2}})\tilde{\Omega}+\frac{1}{C'^2_{2}}-\frac{1}{C'^2_{2}}\frac{1}{\tilde{\Omega}}\rb,\nonumber \\
&& ~~~~~~~~~~~~~~~~~~~~~~~~\text{for}~\frac{2}{C'_{2}-1}>\tilde{\Omega}>\frac{2}{1+C'_{2}}\nonumber\\
&& =\frac{2}{C'^2_{2}},\hspace{1.7cm}\text{for}~\tilde{\Omega}>\frac{2}{C'_{2}-1}
\label{Im-ACsigmaXY-c2>1}
\eea
\begin{figure}[H]
\begin{center}
\includegraphics[width=2.5in,height=3.2in, angle=270]{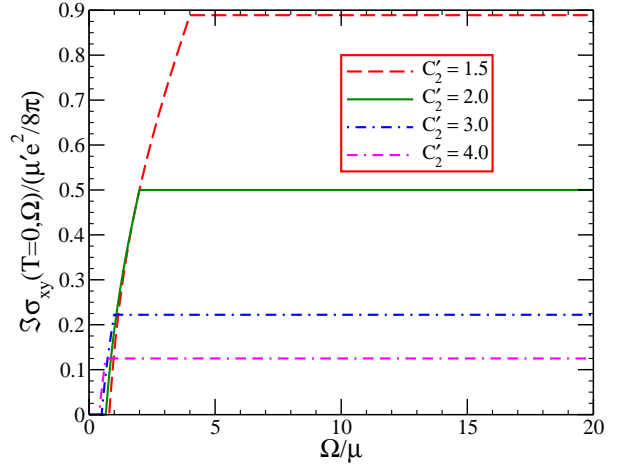} 
\vspace{0.5cm}
\caption{(Color online) Imaginary anomalous Hall conductivity $\Im\sigma_{xy}(T=0,\Omega)$ in units of $\frac{\mu' e^2}{8\pi}$ (where $\mu'=\mu/v$) is plotted against the 
photon energy 
$\Omega$ normalized by $\mu$ for four different values of the tilt parameter $C'_{2}$ which represent the overtilted case or type-II WSM. Here 
$\Im\sigma_{xy}(T=0,\Omega)$ is described by the same functional form as in the type-I WSM case in the range $\frac{2}{1+C'_{2}}<\frac{\Omega}{\mu}<\frac{2}{C'_{2}-1}$. 
But unlike type-I WSM it acquires some finite constant value $\frac{2}{C'^2_{2}}$ which is independent of $\Omega$ for $\frac{\Omega}{\mu}>\frac{2}{C'_{2}-1}$ as 
described in Eq.(\ref{Im-ACsigmaXY-c2>1}).} 
\label{xy-c2>1}
\end{center}
\end{figure}

In Fig.(\ref{xy-0<c2<1}) we show our result for the imaginary part of the finite frequency ($\Omega$) anomalous Hall conductivity $\Im\sigma_{xy}(T=0,\Omega)$ at zero 
temperature $T=0$ in units of $\frac{\mu' e^2}{8\pi}$ as a function of $\Omega/\mu$. Here $\mu'$ means $\mu/v$. The chemical potential scales out of these curves. Results 
for three values of 
$C'_{2}$ are shown namely $C'_{2}=0.1$ (solid green), $C'_{2}=0.5$ (dashed red) and $C'_{2}=0.9$(dash-dotted blue). In all three cases the Hall conductivity is non zero 
only in the photon energy range  $\frac{2}{1+C'_{2}}<\frac{\Omega}{\mu}<\frac{2}{1-C'_{2}}$ for $C'_{2}<1$. These results are to be contrasted with those for $C'_{2}>1$ 
(overtilted) which are presented in Fig.(\ref{xy-c2>1}). Here five values of $C'_{2}$ are shown. The dashed red curve is for $C'_{2}=1.5$, the solid green for $C'_{2}=2.0$, 
the dash-dotted blue for $C'_{2}=3.0$ and double-dashed-dotted purple for $C'_{2}=4.0$. Now $\Im(\sigma_{xy}(T=0,\Omega))$ is still zero for 
$\frac{\Omega}{\mu}<\frac{2}{C'_{2}+1}$ but has a similar functional dependence in a slightly different range $\frac{2}{1+C'_{2}}<\frac{\Omega}{\mu}<\frac{2}{C'_{2}-1}$
than in Fig.(\ref{xy-0<c2<1}) and more importantly $\Im\sigma_{xy}(T=0,\Omega)$ is not zero for $\frac{\Omega}{\mu}>\frac{2}{C'_{2}-1}$, rather it takes on a constant 
value which depends only on the size of $C'_{2}$.

\section{A.C. conductivity for right and left handed polarization}
\label{sec:III}

Now we work on the dynamic diagonal optical conductivity in the same spirit as for the anomalous conductivity. The dynamic diagonal conductivity $\sigma_{xx}(\Omega)$ is 
defined in the same way from the current-current correlation $\Pi_{xx}(\Omega,\q)$ and we get the following form for $\sigma_{xx}(T,\Omega)$,
\bea
&& \sigma_{xx}(T,\Omega)=-\frac{e^2 v^3}{2\pi^2 \Omega} \sum_{s'=\pm}\int^{\Lambda-s'Q}_{-\Lambda-s'Q} dk_{z} \int^{\infty}_{0} \frac{k_{\perp}dk_{\perp}}{k} \nonumber\\
&& \{f(C_{s'}k_{z}+vk)-f(C_{s'}k_{z}-vk)\} \lp2k^2_{z}+ k^2_{\perp}\rp \nonumber\\
&& \lb \pi\delta(4v^2k^2-\Omega^2)-\frac{\imath}{(4v^2k^2-\Omega^2)}\rb. 
\eea
The real part $\Re {\sigma_{xx}}(T,\Omega)$ is the absorptive part and can be written as,
\bea
&& \Re {\sigma_{xx}(T,\Omega)}=-\frac{e^2 v^2}{8\pi \Omega^2}\sum_{s'=\pm}\int^{\Omega/2v}_{-\Omega/2v} dk_{z} \nonumber \\
&& \{f(C_{s'}k_{z}+\frac{\Omega}{2})-f(C_{s'}k_{z}-\frac{\Omega}{2})\} \lp k^2_{z}+ \frac{\Omega^2}{4v^2}\rp.
\eea
It has already been worked out in Ref.[\onlinecite{Carbotte1}]. 

\begin{figure}
\hspace{-1.0cm}
\includegraphics[width=3.5cm,height=4.2cm,angle=270]{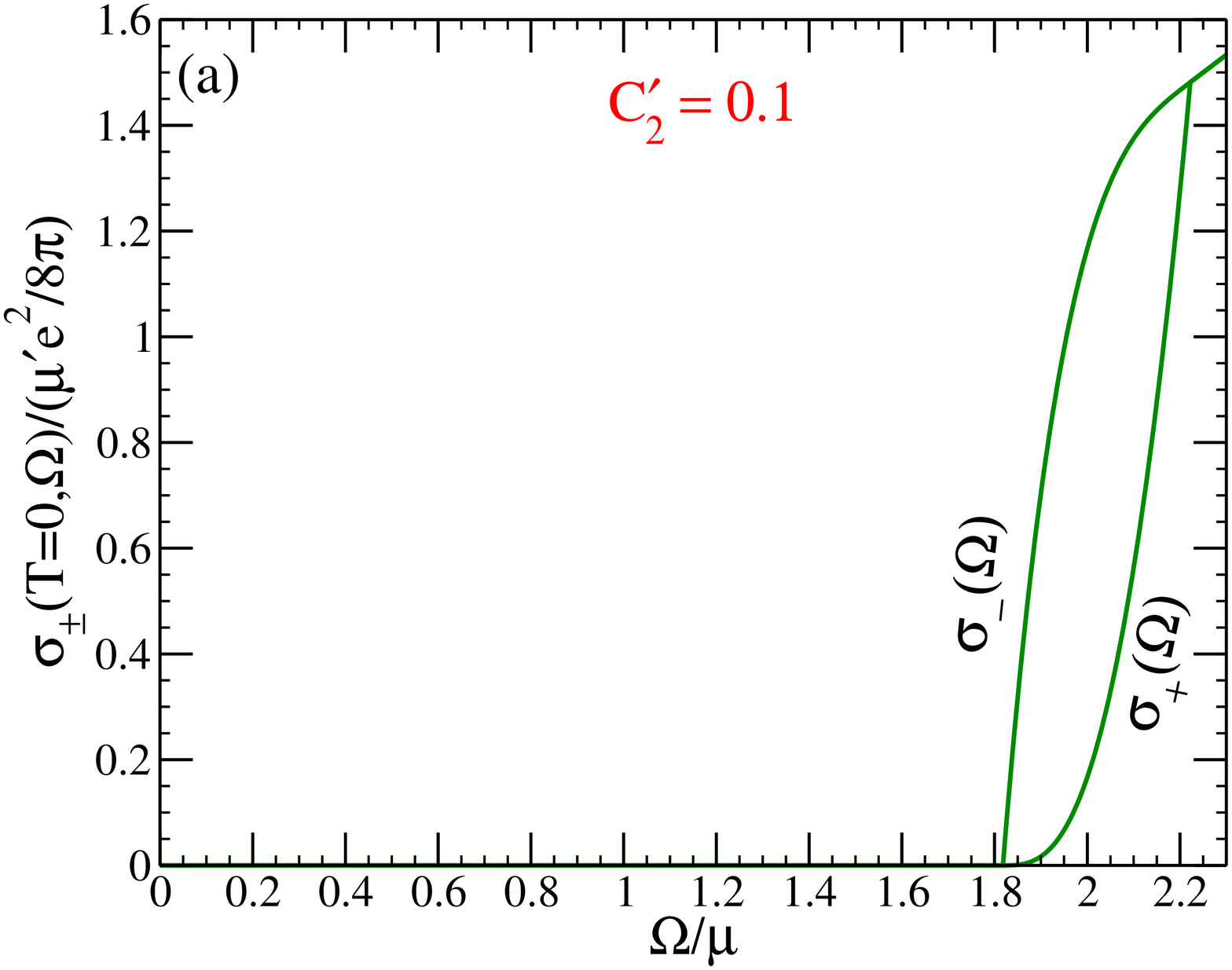}
\includegraphics[width=3.5cm,height=4.2cm,angle=270]{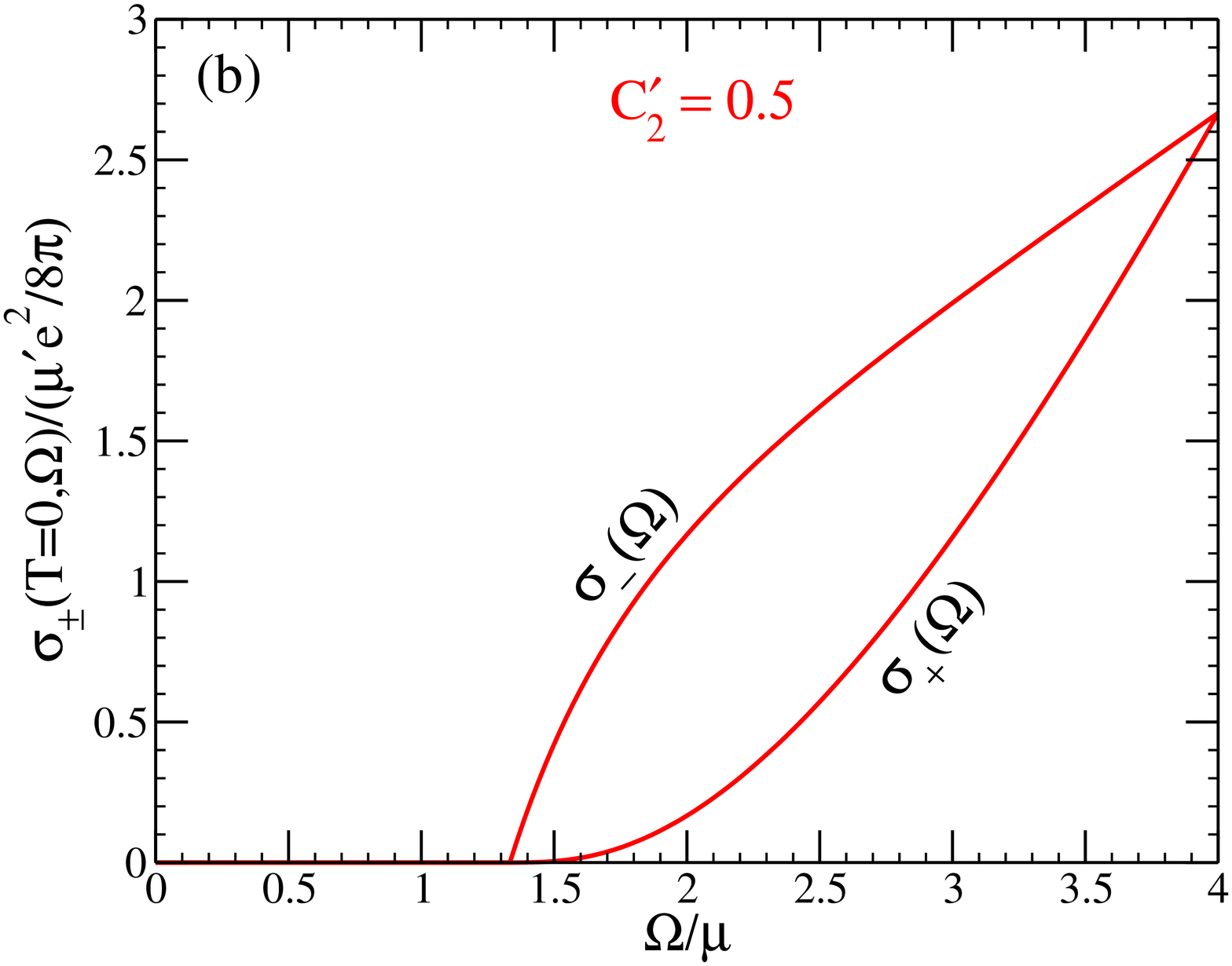}
\includegraphics[width=3.5cm,height=4.2cm,angle=270]{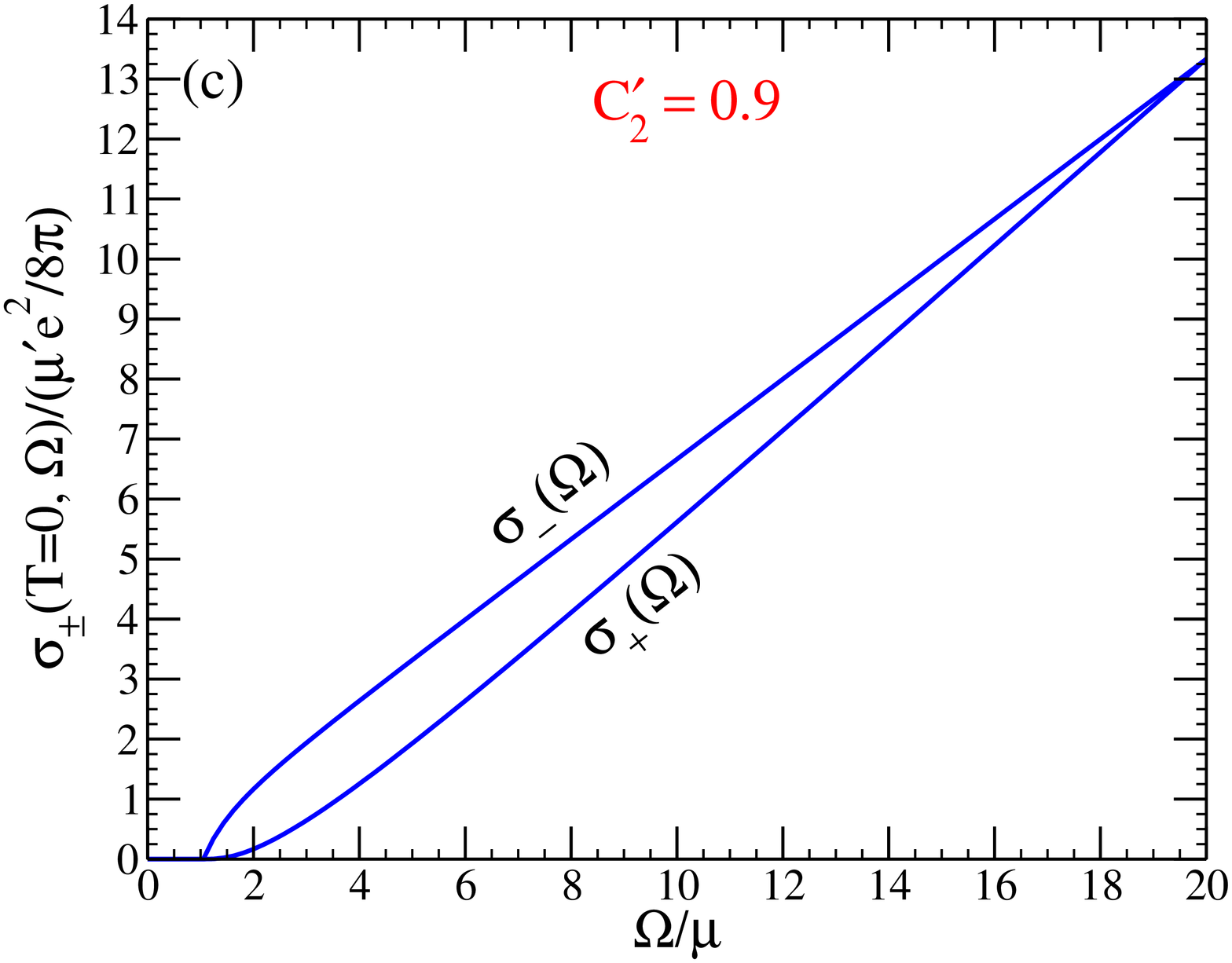}
\caption{(Color online) Here we show the variation of both $\sigma_{+}(T=0,\Omega)$ and $\sigma_{-}(T=0,\Omega)$ in the units of $\frac{\mu' e^2}{8\pi}$ (where $\mu'=\mu/v$) against the 
variation of $\Omega/\mu$ for three representative values of $C'_{2}$ namely, (a)$C'_{2}=0.1$, (b)$C'_{2}=0.5$ and (c) $C'_{2}=0.9$. As shown in Eq.(\ref{ACsigmaP-0<c2<1}) and 
(\ref{ACsigmaM-0<c2<1}) both $\sigma_{+}(T=0,\Omega)$ and $\sigma_{-}(T=0,\Omega)$ is zero below $\frac{\Omega}{\mu}=\frac{2}{1+C'_{2}}$. In the intermediate range
$\frac{2}{1-C'_{2}}>\frac{\Omega}{\mu}>\frac{2}{1+C'_{2}}$ both of them vary differently with $\Omega/\mu$ in such a way that $\sigma_{-}(T=0,\Omega)$ is always greater than
$\sigma_{+}(T=0,\Omega)$ and together they form a `leaf'-like structure which vary in shape or thickness with the varying amount of tilt $C'_{2}$. We consider it as a very 
important signature for WSM type-I materials with two oppositely tilted cones and can be probed experimentally. Beyond this range of $\Omega/\mu$ both $\sigma_{+}(T=0,\Omega)$ 
and $\sigma_{-}(T=0,\Omega)$ merge together into a single straight line, independent of $C'_{2}$.}
\label{sigmaPM-0<c2<1}
\end{figure}

In our notation we get for $0<C'_{2}<1$ (WSM type-I case),
\bea
&& \frac{\Re\sigma_{xx}(T=0,\Omega)}{\mu' e^2/8\pi}=0,\hspace{0.3cm}\text{for}~\tilde{\Omega}<\frac{2}{1+C'_{2}}\nonumber\\
&& \hspace{-0.6cm}=\frac{1}{12}(4+\frac{3}{C'_{2}}+\frac{1}{C'^3_{2}})\tilde{\Omega}-\frac{1}{2}(\frac{1}{C'_{2}}+\frac{1}{C'^3_{2}})+\frac{1}{C'^3_{2}\tilde{\Omega}}- \frac{2}{3C'^3_{2}\tilde{\Omega}^2},\nonumber\\
&& \hspace{3.5cm}\text{for}~\frac{2}{1-C'_{2}}>\tilde{\Omega}>\frac{2}{1+C'_{2}}\nonumber\\
&& =\frac{2\tilde{\Omega}}{3},\hspace{2.5cm}\text{for}~\tilde{\Omega}>\frac{2}{1-C'_{2}}
\label{Re-ACsigmaXX-0<c2<1}
\eea
For $C'_{2}>1$ (overtilted WSM type-II case) we get 
\bea
&& \frac{\Re\sigma_{xx}(T=0,\Omega)}{\mu' e^2/8\pi}=0,\hspace{0.3cm}\text{for}~~~~ \tilde{\Omega}<\frac{2}{1+C'_{2}}\nonumber\\
&& \hspace{-0.6cm}=\frac{1}{12}(4+\frac{3}{C'_{2}}+\frac{1}{C'^3_{2}})\tilde{\Omega}-\frac{1}{2}(\frac{1}{C'_{2}}+\frac{1}{C'^3_{2}})+\frac{1}{C'^3_{2}\tilde{\Omega}}- \frac{2}{3C'^3_{2}\tilde{\Omega}^2},\nonumber\\
&& \hspace{3.5cm}\text{for}~\frac{2}{C'_{2}-1}>\tilde{\Omega}>\frac{2}{1+C'_{2}}\nonumber\\
&&=\frac{1}{6}(\frac{3}{C'_{2}}+\frac{1}{C'^3_{2}})\tilde{\Omega}+ \frac{2}{C'^3_{2}\tilde{\Omega}},\hspace{0.2cm}\text{for}~\tilde{\Omega}>\frac{2}{C'_{2}-1}
\label{Re-ACsigmaXX-c2>1}
\eea

We can construct from Eq.(\ref{Im-ACsigmaXY-0<c2<1}) and (\ref{Im-ACsigmaXY-c2>1}) for $\Im\sigma_{xy}(T=0,\Omega)$ and Eq.(\ref{Re-ACsigmaXX-0<c2<1}) and 
(\ref{Re-ACsigmaXX-c2>1}) for $\Re\sigma_{xx}(T=0,\Omega)$ the absorptive part of the conductivity associated with polarized light namely for right and left polarization, 
\be
\sigma_{\pm}(T=0,\Omega)= \Re\sigma_{xx}(T=0,\Omega)\mp \Im\sigma_{xy}(T=0,\Omega) .
\label{sigmaPM}
\ee
Here we will stick to the assumption that $C_{1}=-C_{2}$ (i.e.$C_{2}$ is assumed to be positive and $C_{1}$ is negative). Assuming $C_{2}=-C_{1}$ (i.e. $C_{1}$ is positive 
and $C_{2}$ negative) merely changes the sign of $\Im\sigma_{xy}(T=0,\Omega)$ in Eq.(\ref{sigmaPM}) which reverses the role of $\sigma_{+}(T=0,\Omega)$ and 
$\sigma_{-}(T=0,\Omega)$. This does not affect the features we will describe in the remaining part. 

\begin{figure}
\includegraphics[width=3.5cm,height=4.2cm, angle=270]{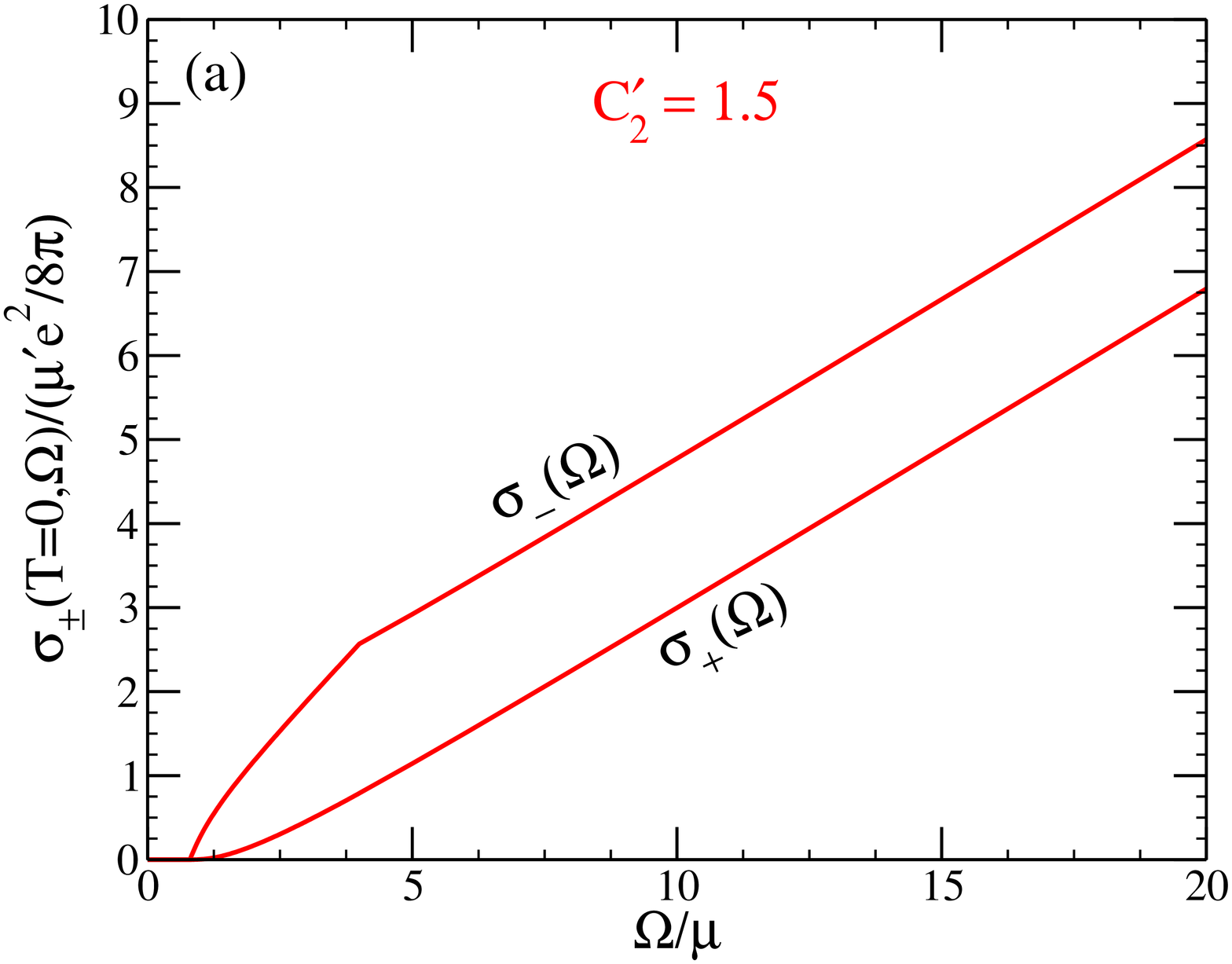}
\includegraphics[width=3.5cm,height=4.2cm, angle=270]{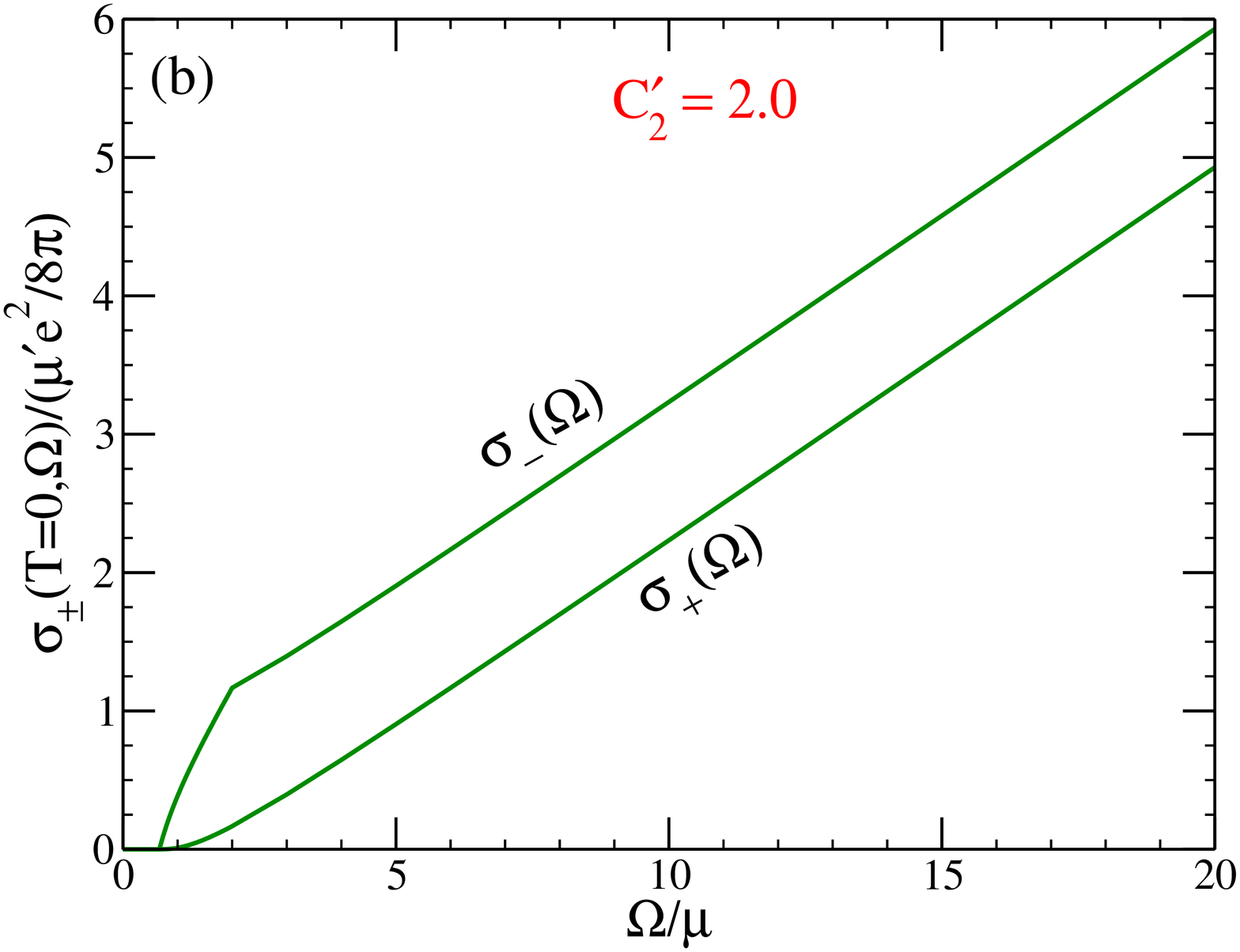}
\includegraphics[width=3.5cm,height=4.2cm, angle=270]{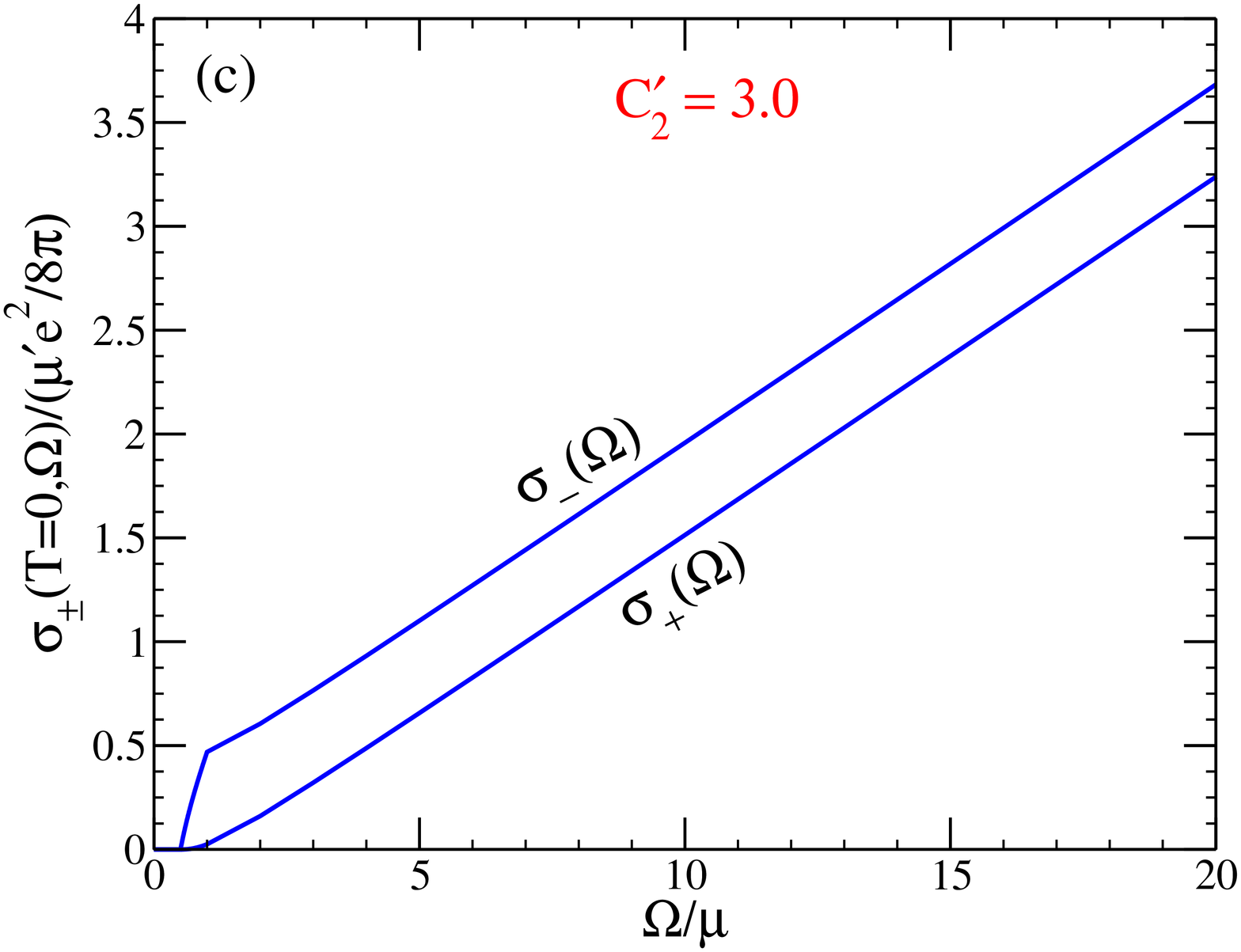}
\includegraphics[width=3.5cm,height=4.2cm, angle=270]{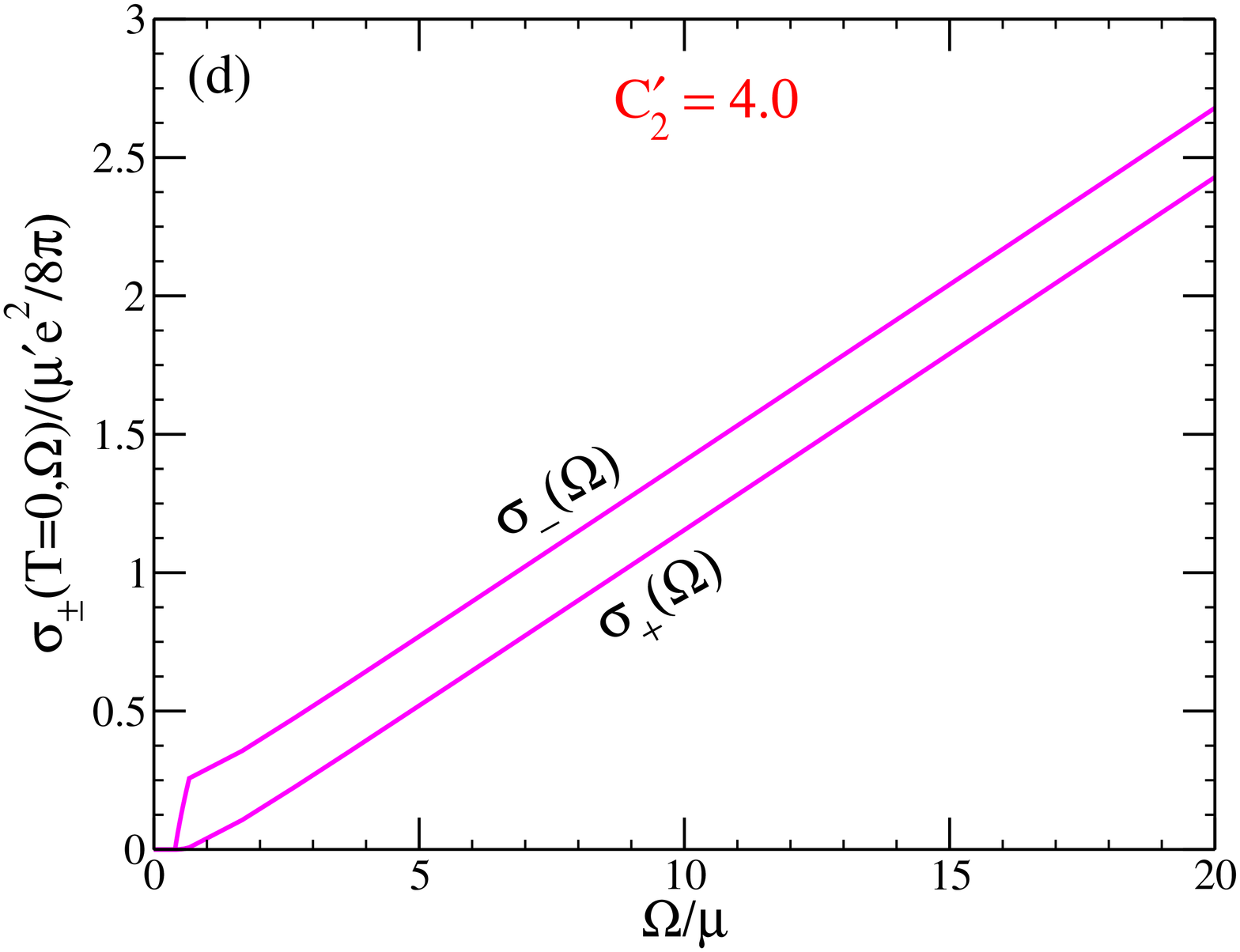}
\caption{(Color online) Here we show the variation of both $\sigma_{+}(T=0,\Omega)$ and $\sigma_{-}(T=0,\Omega)$ in the units of $\frac{\mu' e^2}{8\pi}$ (where $\mu'=\mu/v$) 
against the variation of
$\Omega/\mu$ for four representative values of $C'_{2}$ namely, (a)$C'_{2}=1.5$, (b)$C'_{2}=2.0$, (c) $C'_{2}=3.0$ and (d)$C'_{2}=4.0$. As shown in 
Eq.(\ref{ACsigmaP-c2>1}) and (\ref{ACsigmaM-c2>1}) both $\sigma_{+}(T=0,\Omega)$ and $\sigma_{-}(T=0,\Omega)$ is zero below $\frac{\Omega}{\mu}=\frac{2}{1+C'_{2}}$. In the 
intermediate range $\frac{2}{C'_{2}-1}>\frac{\Omega}{\mu}>\frac{2}{1+C'_{2}}$ both of them vary differently with $\Omega/\mu$ in such a way that $\sigma_{-}(T=0,\Omega)$ is 
always greater than $\sigma_{+}(T=0,\Omega)$. Beyond this range of $\Omega/\mu$,  $\sigma_{+}(T=0,\Omega)$ and $\sigma_{-}(T=0,\Omega)$ are parallel to each other by an amount 
$\frac{4}{C'^2_{2}}$ (independent of $\Omega/\mu$). This characteristic is very special to WSM type-II materials for which the tilt is greater than one  and the two cones 
are oppositely tilted.}
\label{sigmaPM-c2>1}
\end{figure}

For $0<C'_{2}<1$ (WSM type-I case) we get,
\bea
&& \frac{\sigma_{+}(T=0,\Omega)}{\mu'e^2/8\pi}=0,~~\text{for}~~~~ \tilde{\Omega}<\frac{2}{1+C'_{2}}\nonumber\\
&& \hspace{-0.6cm}=\frac{1}{12}(1+\frac{3}{C'_{2}}+\frac{3}{C'^2_{2}}+\frac{1}{C'^3_{2}})\tilde{\Omega}-\frac{1}{2}(\frac{1}{C'_{2}}+\frac{2}{C'^2_{2}}+ \frac{1}{C'^3_{2}})+\nonumber\\
&& (\frac{1}{C'^2_{2}}+\frac{1}{C'^3_{2}})\frac{1}{\tilde{\Omega}}- \frac{2}{3C'^3_{2}\tilde{\Omega}^2},\nonumber\\
&&~~~~~~~~~~~~~~~~~~~~~\text{for}~ \frac{2}{1-C'_{2}}>\tilde{\Omega}>\frac{2}{1+C'_{2}}\nonumber\\
&& =\frac{2\tilde{\Omega}}{3},\hspace{1.5cm}\text{for}~ \tilde{\Omega}>\frac{2}{1-C'_{2}}
\label{ACsigmaP-0<c2<1}
\eea

and 

\bea
&& \frac{\sigma_{-}(T=0,\Omega)}{\mu'e^2/8\pi}=0,~~\text{for}~~~~ \tilde{\Omega}<\frac{2}{1+C'_{2}}\nonumber\\
&& \hspace{-0.6cm}=\frac{1}{12}(7+\frac{3}{C'_{2}}-\frac{3}{C'^2_{2}}+\frac{1}{C'^3_{2}})\tilde{\Omega}-\frac{1}{2}(\frac{1}{C'_{2}}-\frac{2}{C'^2_{2}}+\frac{1}{C'^3_{2}})-\nonumber\\
&& (\frac{1}{C'^2_{2}}-\frac{1}{C'^3_{2}})\frac{1}{\tilde{\Omega}}- \frac{2}{3C'^3_{2}\tilde{\Omega}^2},\nonumber\\
&& ~~~~~~~~~~~~~~~~~~~~~~\text{for}~ \frac{2}{1-C'_{2}}>\tilde{\Omega}>\frac{2}{1+C'_{2}}\nonumber\\
&& =\frac{2\tilde{\Omega}}{3},\hspace{1.5cm}\text{for}~ \tilde{\Omega}>\frac{2}{1-C'_{2}}
\label{ACsigmaM-0<c2<1}
\eea

For $C'_{2}>1$ (overtilted WSM type-II case),
\bea
&& \frac{\sigma_{+}(T=0,\Omega)}{\mu'e^2/8\pi}=0,~~\text{for}~~~~ \tilde{\Omega}<\frac{2}{1+C'_{2}}\nonumber\\
&& \hspace{-0.6cm}=\frac{1}{12}(1+\frac{3}{C'_{2}}+\frac{3}{C'^2_{2}}+\frac{1}{C'^3_{2}})\tilde{\Omega}-\frac{1}{2}(\frac{1}{C'_{2}}+\frac{2}{C'^2_{2}}+\frac{1}{C'^3_{2}})+\nonumber\\ 
&& (\frac{1}{C'^2_{2}}+\frac{1}{C'^3_{2}})\frac{1}{\tilde{\Omega}}- \frac{2}{3C'^3_{2}\tilde{\Omega}^2},\nonumber\\
&&~~~~~~~~~~~~~~~~~~~~~~~~~~~\text{for}~\frac{2}{C'_{2}-1}>\tilde{\Omega}>\frac{2}{1+C'_{2}}\nonumber\\
&& \hspace{-0.2cm}=\frac{1}{6}(\frac{3}{C'_{2}}+\frac{1}{C'^3_{2}})\tilde{\Omega}-\frac{2}{C'^2_{2}}+\frac{2}{C'^3_{2}\tilde{\Omega}},~\text{for}~\tilde{\Omega}>\frac{2}{C'_{2}-1}\nonumber\\
&&
\label{ACsigmaP-c2>1}
\eea
and 
\bea
&& \frac{\sigma_{-}(T=0,\Omega)}{\mu'e^2/8\pi}=0,~~\text{for}~~~~ \tilde{\Omega}<\frac{2}{1+C'_{2}}\nonumber\\
&&  \hspace{-0.6cm}=\frac{1}{12}(7+\frac{3}{C'_{2}}-\frac{3}{C'^2_{2}}+\frac{1}{C'^3_{2}})\tilde{\Omega}-\frac{1}{2}(\frac{1}{C'_{2}}-\frac{2}{C'^2_{2}}+\frac{1}{C'^3_{2}})-\nonumber \\
&& (\frac{1}{C'^2_{2}}-\frac{1}{C'^3_{2}})\frac{1}{\tilde{\Omega}}- \frac{2}{3C'^3_{2}\tilde{\Omega}^2},\nonumber\\
&& ~~~~~~~~~~~~~~~~~~~~~~~~~~~\text{for}~~~~\frac{2}{C'_{2}-1}>\tilde{\Omega}>\frac{2}{1+C'_{2}}\nonumber\\
&& \hspace{-0.2cm}=\frac{1}{6}(\frac{3}{C'_{2}}+\frac{1}{C'^3_{2}})\tilde{\Omega}+\frac{2}{C'^2_{2}}+\frac{2}{C'^3_{2}\tilde{\Omega}},~\text{for}~~~~\tilde{\Omega}>\frac{2}{C'_{2}-1}\nonumber\\
&&
\label{ACsigmaM-c2>1}
\eea

Results for $\sigma_{\pm}(T=0,\Omega)$ based on Eq.(\ref{ACsigmaP-0<c2<1}) and (\ref{ACsigmaM-0<c2<1}) for $0<C'_{2}<1$  are presented in Fig.(\ref{sigmaPM-0<c2<1}) while 
results for the case $C'_{2}>1$ (overtilted) based on Eq.(\ref{ACsigmaP-c2>1}) and (\ref{ACsigmaM-c2>1}) are shown in Fig.(\ref{sigmaPM-c2>1}). These two regime show quite 
distinct behaviors. In both figures $\sigma_{\pm}(T=0,\Omega)$ is presented in units of $\frac{\mu' e^2}{8\pi}$ (where $\mu'=\mu/v$) as a function of photon energy 
$\Omega$ also normalized to the 
chemical potential $\mu$. $\sigma_{+}(T=0,\Omega)$ and $\sigma_{-}(T=0,\Omega)$ are compared and results for three values of $C'_{2}$ are shown. $C'_{2}=0.1$ by green curve (upper left frame),
$C'_{2}=0.5$ by red curve (upper right frame) and $C'_{2}=0.9$ by blue curve (bottom frame). The range of photon energies for which $\sigma_{+}(T=0,\Omega)$ and $\sigma_{-}(T=0,\Omega)$ are 
non-zero is of course restricted by the range for which $\sigma_{xy}(T=0,\Omega)$ is non-zero as shown in Fig(\ref{xy-0<c2<1}) which applies to the case $0<C'_{2}<1$. As
$C'_{2}$ is increased the range of interest expands both to lower and to higher energies $\frac{\Omega}{\mu}$ with the upper limit getting even longer as $C'_{2}$ 
approaches one at which point $\frac{1}{(1-C'_{2})}$ tends towards infinity and $\sigma_{\pm}(T=0,\Omega)$ will remain finite to high energies. Note that $\sigma_{+}(T=0,\Omega)$ is
always smaller than $\sigma_{-}(T=0,\Omega)$. For the overtilted case $C'_{2}>1$ the behavior of $\sigma_{+}(T=0,\Omega)$ and $\sigma_{-}(T=0,\Omega)$ are shown in Fig.(\ref{sigmaPM-c2>1}) is very different 
from that in Fig.(\ref{xy-0<c2<1}). In particular there is now a large range of $\Omega$ over which  $\sigma_{+}(T=0,\Omega)$ and $\sigma_{-}(T=0,\Omega)$ are parallel to each other. That this 
is so can be seen from our analytic results (\ref{ACsigmaP-c2>1}) and (\ref{ACsigmaM-c2>1}). For $\frac{\Omega}{\mu}>\frac{2}{C'_{2}-1}$ only the constant term 
$\frac{2}{C'^2_{2}}$ is different. It appears with a plus sign in Eq.(\ref{ACsigmaM-c2>1}) while its sign is negative in Eq.(\ref{ACsigmaP-c2>1}).

\section{The Hall angle as a function of photon energy}
\label{sec:IV}

The Hall angle $\theta_{H}(T=0,\Omega)$ as a function of photon energy $\Omega$ is defined as,
\be
\theta_H(T=0,\Omega)=\frac{\Re\sigma_{+}(\Omega)-\Re\sigma_{-}(\Omega)}{\Re\sigma_{+}(\Omega)+\Re\sigma_{-}(\Omega)}= -\frac{\Im\sigma_{xy}(\Omega)}{\Re\sigma_{xx}(\Omega)}
\label{Hall-angle}
\ee

For $0<C'_{2}<1$ (WSM type-I case),
\bea
&& \theta_H(T=0,\Omega)=0,\hspace{0.7cm}\text{for}~\tilde{\Omega}<\frac{2}{1+C'_{2}}\nonumber\\
&& =-\frac{3C'_{2}\{(C'^2_{2}-1)\tilde{\Omega}^3+4\tilde{\Omega}^2-4\tilde{\Omega}\}}{(4C'^3_{2}+3C'^2_{2}+1)\tilde{\Omega}^3-6(C'^2_{2}+1)\tilde{\Omega}^2+12\tilde{\Omega}-8},\nonumber\\
&& \hspace{3.5cm}\text{for}~\frac{2}{1-C'_{2}}>\tilde{\Omega}>\frac{2}{1+C'_{2}}\nonumber\\
&& =0,\hspace{2.8cm}\text{for}~\tilde{\Omega}>\frac{2}{1-C'_{2}}
\label{Hall-angle-WSMI}
\eea
At $\frac{\Omega}{\mu}=\frac{2}{1+C'_{2}}$ the algebraic expression in Eq.(\ref{Hall-angle-WSMI}) reduces to one.
\begin{figure}[H]
\begin{center}
\includegraphics[width=2.5in,height=3.2in, angle=270]{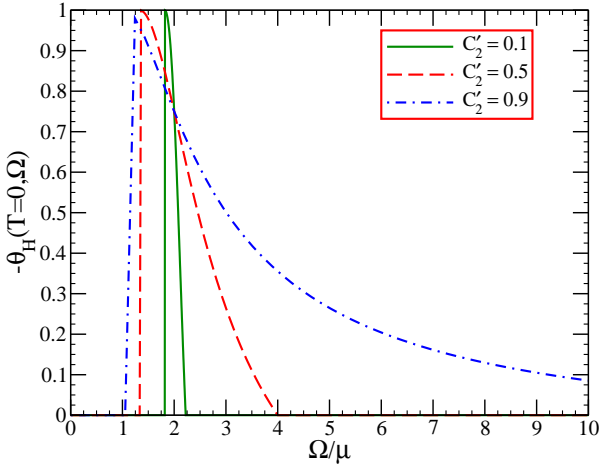} 
\vspace{0.5cm}
\caption{(Color online)Here we show the variation of the negative of the Hall angle $\theta_{H}(T=0,\Omega)$ in radians against the variation of $\Omega/\mu$ for three 
representative values of $C'_{2}$ namely $0.1, 0.5$ and $0.9$ specific to the WSM type-I case. It is only non-zero in the range 
$\frac{2}{1-C'_{2}}>\tilde{\Omega}>\frac{2}{1+C'_{2}}$ as described in Eq.(\ref{Hall-angle-WSMI}). Both below and above this range they goes to zero.} 
\label{HA-0<c2<1}
\end{center}
\end{figure}

For $C'_{2}>1$ (overtilted WSM type-II case) we have,
\bea
&& \theta_H(T=0,\Omega)=0,\hspace{0.7cm} \text{for}~\tilde{\Omega}<\frac{2}{1+C'_{2}}\nonumber\\
&& =-\frac{3C'_{2}\{(C'^2_{2}-1)\tilde{\Omega}^3+4\tilde{\Omega}^2-4\tilde{\Omega}\}}{(4C'^3_{2}+3C'^2_{2}+1)\tilde{\Omega}^3-6(C'^2_{2}+1)\tilde{\Omega}^2+12\tilde{\Omega}-8},\nonumber\\
&& \hspace{3.5cm}\text{for}~\frac{2}{C'_{2}-1}>\tilde{\Omega}>\frac{2}{1+C'_{2}}\nonumber\\
&& =-\frac{12 C'_{2}\tilde{\Omega}}{(3C'^2_{2}+1)\tilde{\Omega}^2+12},\hspace{0.2cm}\text{for}~\tilde{\Omega}>\frac{2}{C'_{2}-1}
\label{Hall-angle-WSMII}
\eea
Again $\theta_H(T=0,\Omega)=1$ at $\frac{\Omega}{\mu}=\frac{2}{1+C'_{2}}$ and for $\frac{\Omega}{\mu}=\infty$ we get,
\be
\theta_H(T=0,\Omega)=-\frac{12 C'_{2}}{(3C'^2_{2}+1)\tilde{\Omega}},
\ee
so that in this case $\theta_H(T=0,\Omega)$ remains finite above $\frac{\Omega}{\mu}=\frac{2}{1+C'_{2}}$ and decays as $\sim \frac{1}{\tilde{\Omega}}$ while for type-I 
$\theta_H(T=0,\Omega)$ is zero above $\frac{\Omega}{\mu}=\frac{2}{C'_{2}-1}$.
\begin{figure}[H]
\begin{center}
\includegraphics[width=2.5in,height=3.2in, angle=270]{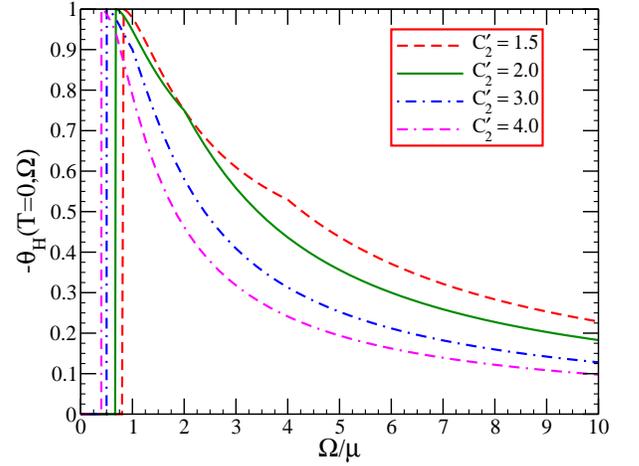} 
\vspace{0.5cm}
\caption{(Color online) We show the variation of the negative of the Hall angle $\theta_{H}(T=0,\Omega)$ in radians against the variation of $\Omega/\mu$ for four 
representative values of $C'_{2}$ namely $1.5, 2.0, 3.0$ and $4.0$ specific to the WSM type-II case. We see that Hall angle is zero only in the range 
$\frac{\Omega}{\mu}<\frac{2}{1+C'_{2}}$. It has the same functional dependence as in the WSM type-I case in the range $\frac{2}{C'_{2}-1}>\tilde{\Omega}>\frac{2}{1+C'_{2}}$ 
as described in Eq.(\ref{Hall-angle-WSMII}). Above this range it decays as $\sim \frac{1}{\tilde{\Omega}}$.} 
\label{HA-c2>1}
\end{center}
\end{figure}

Our result based on the simple algebraic expressions Eq.(\ref{Hall-angle-WSMI}) for type-I and Eq.(\ref{Hall-angle-WSMII}) for type-II are shown in Fig.(\ref{HA-0<c2<1}) 
and Fig.(\ref{HA-c2>1}) respectively. The same three values of $C'_{2}$ that we used in previous sections are shown as solid green line ($C'_{2}=0.1$), dashed red 
($C'_{2}=0.5$) and dash-dotted blue ($C'_{2}=0.9$) in Fig.(\ref{HA-0<c2<1}) for the type-I case. Note how the range of photon energies for which $\theta_H(T=0,\Omega)$ is 
finite increases as $C'_{2}$ increases. For the type-II case we present in Fig.(\ref{HA-c2>1}) results for four values of the tilt namely $C'_{2}=1.5$ dashed red curve, 
$C'_{2}=2.0$ solid green, $C'_{2}=3.0$ dash-dotted blue and $C'_{2}=4.0$ double-dashed dotted purple. For this case the Hall angle remains finite for all 
$\frac{\Omega}{\mu}>\frac{1}{(1+C'_{2})}$ although it becomes small as $\frac{\Omega}{\mu}$ becomes large.

\section{Summary and conclusions}
\label{sec:V}

We find that the dynamic anomalous Hall conductivity $\sigma_{xy}(T=0,\Omega)$ normalized to the chemical potential $\mu$ in units of $\frac{e^2}{8\pi v}$ ($v$ the Fermi 
velocity) as a function of photon energy $\Omega$ normalized to $\mu$ is a universal function dependent only on the tilt of the Dirac cone. For a pair of Weyl nodes 
oppositely tilted and of opposite chirality the absorptive part of the Hall conductivity $\Im\sigma_{xy}(T=0,\Omega)$ in type-I WSM is non-zero only in a finite interval 
of photon energies, $\frac{2}{1-C'_{2}}>\frac{\Omega}{\mu}>\frac{2}{1+C'_{2}}$. In sharp contrast for type-II there is no upper bound on $\Omega$. The 
$\Im\sigma_{xy}(T=0,\Omega)$ remains zero below $\frac{2}{1+C'_{2}}$, rises sharply in the interval $\frac{2}{1-C'_{2}}>\frac{\Omega}{\mu}>\frac{2}{1+C'_{2}}$ and becomes 
constant equal to $\frac{2}{C'^2_{2}}$ above $\frac{\Omega}{\mu}=\frac{2}{C'_{2}-1}$. This can be taken as a signature for overtilting.

The absorptive part of the A.C. optical conductivity associated with right and left handed polarized light $\sigma_{\pm}$  in units of $\frac{\mu e^2}{8\pi v}$ as function 
of $\frac{\Omega}{\mu}$ is again a universal function dependent only on the tilt $C'_{2}$ and $\frac{\Omega}{\mu}$, and is given by a specific algebraic expressions
(Eq.(\ref{ACsigmaP-0<c2<1}) and (\ref{ACsigmaM-0<c2<1})). For a type-I WSM $\sigma_{+}(T=0,\Omega)$ and $\sigma_{-}(T=0,\Omega)$ differ from each other only in the interval 
$\frac{2}{1+C'_{2}}<\frac{\Omega}{\mu}<\frac{2}{1-C'_{2}}$ with $\sigma_{+}(T=0,\Omega)$ always smaller than $\sigma_{-}(T=0,\Omega)$ except at the boundaries where they are 
equal. Both $\sigma_{\pm}(T=0,\Omega)$ are zero for $\frac{\Omega}{\mu}<\frac{2}{1+C'_{2}}$ and for $\frac{\Omega}{\mu}>\frac{2}{1-C'_{2}}$ they both reduce to the same 
value equal to $\Re\sigma_{xx}(T=0,\Omega)$ because in the interval the anomalous Hall conductivity is zero. In the overtilted regime (type-II WSM) $\sigma_{\pm}(T=0,\Omega)$ 
behaves very differently than for the type-I case. There still exists a lower frequency $\frac{\Omega}{\mu}=\frac{2}{1+C'_{2}}$ below which right and left hand optical 
response is zero. This is followed by a frequency range $\frac{2}{1+C'_{2}}<\frac{\Omega}{\mu}<\frac{2}{1-C'_{2}}$ in which $\sigma_{-}(T=0,\Omega)$ rises faster than 
$\sigma_{+}(T=0,\Omega)$. But above $\frac{\Omega}{\mu}=\frac{2}{C'_{2}-1}$ the two curves become parallel to each other displaced by a constant amount to $\frac{4}{C'^2_{2}}$ 
in our chosen units.

We give simple analytic algebraic formulas for the Hall angle $\theta_{H}(T=0,\Omega)$ as a function of the photon energy $\frac{\Omega}{\mu}$. These appear as 
Eq.(\ref{Hall-angle-WSMI}) and (\ref{Hall-angle-WSMII}). The Hall angle is zero for $\frac{\Omega}{\mu}<\frac{2}{1+C'_{2}}$. Just above this photon energy it has value 
one and this value gets reduced as $\frac{\Omega}{\mu}$ is increased. For the type-I WSM case there is an upper photon energy $\frac{\Omega}{\mu}=\frac{2}{1-C'_{2}}$ 
above which the Hall angle is zero. For type-II WSM no such upper photon energy exists and $\theta_{H}(T=0,\Omega)$ remains finite and decays as 
$\frac{12 C'_{2}}{(3C'^2_{2}+1)(\Omega/\mu)}$ as $\Omega/\mu\rightarrow \infty$.

\subsection*{Acknowledgments}
Work supported in part by the Natural Sciences and Engineering Research Council of Canada (NSERC) and by the Canadian Institute for Advanced Research (CIAR).

\end{document}